\documentclass[12pt]{article}
\usepackage{epsf}
\newcommand{\lesssim}{\mathrel{\hbox{\rlap{\hbox{\lower4pt\hbox{$\sim$}}}\hbox{$
<$}}}}
\newcommand{\gtrsim}{\mathrel{\hbox{\rlap{\hbox{\lower4pt\hbox{$\sim$}}}\hbox{$>
$}}}}
\newcommand{\bg}[1]{\mbox{\boldmath $#1$}}

\begin{document}

\thispagestyle{empty}

\begin{centering}
\parindent 0mm

{\large\bf HEATING OF GAS INSIDE RADIO SOURCES TO MILDLY RELATIVISTIC
TEMPERATURES VIA INDUCED COMPTON SCATTERING} 

\vspace{5mm}
{\bf S.Y. Sazonov, R.A. Sunyaev}

\vspace{5mm}
{\sl Space Research Institute, Moscow, Russia\\
MPI f\"ur Astrophysik, Garching, Germany}

\end{centering}

\vspace{5mm}
Measured values of the brightness temperature of low-frequency
synchrotron radiation emitted by powerful extragalactic sources reach
$10^{11}$--$10^{12}$~K. If some amount of nonrelativistic ionized gas
is present within such sources, it should be heated as a result of
induced Compton scattering of the radiation. If this heating is
counteracted by cooling due to inverse Compton scattering of the 
same radio radiation, then the plasma can be heated up to mildly relativistic
temperatures $kT\sim 10$--$100$~keV. The stationary electron velocity
distribution can be either relativistic Maxwellian or quasi-Maxwellian
(with the high-velocity tail suppressed), depending on the efficiency
of Coulomb collisions and other relaxation processes. We derive
several easy-to-use approximate expressions for the induced Compton
heating rate of mildly relativistic electrons in an isotropic
radiation field, as well as for the stationary distribution function
and temperature of electrons.

We give analytic expressions for the kernel of the integral kinetic
equation (one as a function of the scattering angle and another for
the case of an isotropic radiation field), which describes the
redistribution of photons in frequency caused by induced Compton
scattering in thermal plasma. These expressions can be used in the
parameter range $h\nu\ll kT\lesssim 0.1mc^2$ (the formulae earlier
published in Sazonov, Sunyaev, 2000 are less accurate). 

%\clearpage

%%%%%%%%%%%%%%%%%%%%%%%%%%%%%%%%%%%%%%%%%%%%%%%%%%%%%%%%%%%%%%%%%%%%%%%%
\section{Introduction}
%%%%%%%%%%%%%%%%%%%%%%%%%%%%%%%%%%%%%%%%%%%%%%%%%%%%%%%%%%%%%%%%%%%%%%%%

Milliarcsecond-resolution interferometric radio observations of the
central parsec-scale regions of active galactic nuclei often
reveal sub-structure emitting low-frequency radiation with inferred
rest-frame brightness temperatures $T_b\sim 10^{11}$--$10^{12}$~K. If
there were thermal plasma present in these regions, it could be heated
efficiently as a result of induced Compton scattering of the radio
radiation on the electrons (Levich, Sunyaev, 1971).

Such a plasma has not been observed so far in extragalactic
sources. On the other hand, there is evidence that relatively 
cold matter constitutes a significant fraction of the total mass
contained in the jets, including their inner ($\le 10^{12}$~cm) regions,
of the famous galactic source SS433 (see Vermeulen, 1992 for a
review). Of course, these jets are only mildly relativistic ($v\sim 0.26
c$) and therefore quite different from the relativistic jets in active
galactic nuclei (which have bulk Lorentz factors $\gamma\sim 5$), but
it is reasonable to expect that some amount of quasi-thermal plasma
may be present in extragalactic jets as well (see, e.g., Celotti et
al., 1998). 

One can derive directly from the Kompaneets kinetic equation the induced
Compton heating rate of thermal electrons located in an isotropic
radiation field (Levich, Sunyaev, 1971). The resulting expression is, however,
only applicable when the electron temperature is relatively low,
$kT\lesssim$ a few keV. As the electrons become more relativistic,
their heating rate monotonically decreases. Theoretical efforts in the
70-s (Vinogradov, Pustovalov, 1972; Blandford, 1973; Blandford,
Scharlemann, 1975) resulted in a number of useful formulae that allow
one to determine the heating rate of relativistic electrons and their
distribution function in various limits, e.g., for ultrarelativistic
electrons, for narrow radiation beams, or for some specific radiation
spectra. Illarionov and Kompaneets (1976) have derived a general
expression that gives the heating rate for an electron moving with an
arbitrary velocity in a given isotropic radiation field. This formula
is, however, rather complex: one needs to compute a 3-dimensional
integral in order to find the heating rate of an ensemble of electrons
with a given velocity distribution (e.g, Maxwellian).

In the present paper we demonstrate that {\sl electrons can be heated
by Compton scattering up to mildly relativistic temperatures
$kT\sim$ a few tens of keV, but not more, in an isotropic radiation
field with $T_b<10^{12}$~K}. For these plasma temperatures, the
nonrelativistic estimates for the heating rate and some relevant
quantities become inaccurate. However, these expressions can be
modified by adding to them a few correction terms, thereby retaining
the original simple structure, as shown below.

It should be noted that although we mentioned above only active galactic
nuclei, the induced Compton effect may play a major role also in other
astrophysical situations. We also note that the problem of plasma
heating near, but outside, a source of low-frequency radiation
requires a special study.

%%%%%%%%%%%%%%%%%%%%%%%%%%%%%%%%%%%%%%%%%%%%%%%%%%%%%%%%%%%%%%%%%%%%%%%%%
\section{Heating and cooling of thermal electrons during Compton scattering}
%%%%%%%%%%%%%%%%%%%%%%%%%%%%%%%%%%%%%%%%%%%%%%%%%%%%%%%%%%%%%%%%%%%%%%%%%

Let us derive the rates of heating and cooling of mildly relativistic
thermal electrons ($kT\lesssim 0.1mc^2$, where $T$ is the electron
temperature) in an isotropic radiation field, as a result of both the
spontaneous and induced Compton scattering processes.

The radiation field may be defined by its spectral energy density,
$\epsilon_\nu$ (measured in units of erg cm$^{-3}$ Hz$^{-1}$), or,
equivalently, by the occupation number in photon phase space,
$n=\epsilon_\nu c^3/(8\pi h\nu^3)$. We require that the spectrum be
sufficiently broad, which means that its characteristic width
$\Delta\nu/\nu$ should be much larger than the typical fractional
frequency shift acquired by a photon during a scattering event 
either due to the Doppler effect, $\sim\pm(kT/mc^2)^{1/2}$, or due to
Compton recoil, $\sim -h\nu/mc^2$. In this case the integral kinetic 
equation describing the time evolution of $\epsilon_\nu$ (caused by
the Compton interaction of the radiation with the electrons) can be
transformed into a Fokker-Planck-type differential equation. In 
the limit of $kT,h\nu\ll mc^2$, the resulting equation is the
Kompaneets (1957) equation.

From the Kompaneets equation one can derive the expressions for the
electron heating and cooling rates (Levich, Sunyaev, 1971), which are fairly
accurate for electron and photon energies $\lesssim 0.01 mc^2$. 
The Kompaneets equation can be extended into the mildly-relativistic
domain ($kT,h\nu\lesssim 0.1mc^2$) by adding to it 
relativistic correction terms $\sim(kT/mc^2)^m (h\nu/mc^2)^n$
(Itoh et al., 1998; Challinor, Lasenby, 1998; Sazonov, Sunyaev,
2000). For example, the first-order generalization of the Kompaneets
equation (which obtains by Fokker-Planck-type expansion of the
integral kinetic equation to the fourth order in $\Delta\nu$) is
\begin{eqnarray}
\frac{\partial n}{\partial t} &=& \frac{\sigma_{\rm T}N_{\rm e}h}{mc} 
\frac{1}{\nu^2}
\frac{\partial}{\partial \nu}\nu^4\left\{n(1+n)+\frac{kT}{h}
\frac{\partial n}{\partial\nu}+\frac{7}{10}\frac{h\nu^2}{mc^2}
\frac{\partial n}{\partial\nu}
\right.
\nonumber\\
&&\left.
+\frac{kT}{mc^2}\left[\frac{5}{2}\left(
n(1+n)+\frac{kT}{h}\frac{\partial n}{\partial\nu}\right)
+\frac{21}{5}\nu \frac{\partial}{\partial\nu}\left(n(1+n)+\frac{kT}{h}
\frac{\partial n} {\partial\nu}\right)
\right.\right.
\nonumber\\
&&\left.\left.
+\frac{7}{10}\nu^2\left(-2\left(\frac{\partial n}{\partial\nu}\right)^2+2(1+2n)
\frac{\partial^2 n}{\partial\nu^2}+\frac{kT}{h}\frac{\partial^3 n}{\partial
\nu^3}\right)\right]\right\},
\label{kgen}
\end{eqnarray}
where $\sigma_{\rm T}$ is the Thomson cross section and $N_{\rm e}$ is
the number density of electrons.

Multiplying Eq.~(\ref{kgen}) by $h\nu$ and integrating over the photon
phase space leads to an expression for the variation of the
radiation total energy density, $\epsilon=\int_0^\infty\epsilon_\nu
d\nu$, and, at the same time, for the change of the mean electron
energy, $E_{\rm e}\approx 1.5 kT(1+1.25 kT/mc^2)$:
\begin{eqnarray}
-\frac{1}{N_{\rm e}}\frac{d\epsilon}{dt}=\frac{dE_{\rm e}}{dt}
=\frac{8\pi\sigma_{\rm T}h}{c^2}
\left[-4\frac{kT}{mc^2}\int_0^\infty\nu^3n d\nu
+\frac{h}{mc^2}\int_0^\infty\nu^4(n+n^2) d\nu
\right.
\nonumber\\
\left.
-10\left(\frac{kT}{mc^2}\right)^2\int_0^\infty\nu^3 n d\nu
-\frac{21}{5}\left(\frac{h}{mc^2}\right)^2\int_0^\infty\nu^5 n d\nu
\right.
\nonumber\\
\left.
+\frac{kT}{mc^2}\frac{h}{mc^2}\left(\frac{47}{2}\int_0^\infty\nu^4(n+n^2)
d\nu
-\frac{21}{5}\int_0^\infty\nu^6\left(\frac{\partial n}
{\partial\nu}\right)^2 d\nu
\right)\right].
\label{transfer}
\end{eqnarray}

The first two terms in the square brackets in Eq.~(\ref{transfer})
describe the electron heating and cooling rates in the
nonrelativistic limit (Peyraud, 1968; Zel'dovich, Levich, 1970; Levich,
Sunyaev, 1971). The additional terms represent relativistic 
corrections of the order of $kT/mc^2$ or $h\nu/mc^2$. Note that the
correction to the heating rate associated with the induced
Compton effect (the corresponding terms are nonlinear in $n$)
includes a term which is expressed through the square of the
derivative, $(\partial n/\partial\nu)^2$. Higher-order [$(kT/mc^2)^2$,
etc.] corrections to the induced-Compton energy transfer rate, which
are not given in Eq.~(\ref{transfer}), depend on higher-order
derivatives of $n$, as is explicitly shown below (in \S 3). 

Using Eq.~(\ref{transfer}), we can, for example, find the energy
transfer rate between electrons and black-body radiation of temperature $T_r$,
\begin{equation}
\frac{d\epsilon}{dt}=4\epsilon\sigma_{\rm T}N_{\rm e}c
\left(\frac{kT}{mc^2}-\frac{kT_r}{mc^2}\right)\left(1
+\frac{5}{2}\frac{kT}{mc^2}-2\pi^2\frac{kT_r}{mc^2}\right).
\label{transfer_planck}
\end{equation}
This expression was originally derived (in a different way) by
Woodward (1970). 

We note that the problem of the energy transfer between
radiation and a thermal plasma via spontaneous Compton scattering is
well studied, and the corresponding part of Eq.~(\ref{transfer})
follows directly from the first moment of the Compton scattering 
kernel (Shestakov et al., 1988; Sazonov, Sunyaev, 2000; also see the
Appendix). We are concerned here with the terms describing the
contribution of induced scattering.

%%%%%%%%%%%%%%%%%%%%%%%%%%%%%%%%%%%%%%%%%%%%%%%%%%%%%%%%%%%%%%%%%%%%%%%%
\section{Induced Compton heating of thermal electrons} 
%%%%%%%%%%%%%%%%%%%%%%%%%%%%%%%%%%%%%%%%%%%%%%%%%%%%%%%%%%%%%%%%%%%%%%%%

Powerful extragalactic radio sources emit low-frequency continuum
radiation that in some cases has a very high brightness temperature, $T_b=n
h\nu/k\sim 10^{11}$--$10^{12}$~K, so that $kT_b>mc^2$. Induced Compton
scattering can be a major mechanism of heating of free electrons
situated in such a radiation field. Moreover, nonrelativistic estimates
lead to the conclusion that electrons can be heated up to relativistic
temperatures $kT\sim$ a fraction of $kT_b$ (Levich, Sunyaev, 1971). Clearly,
nonrelativistic treatment is inappropriate here. Below we show that in
reality, electron temperatures achievable in this situation are only
mildly relativistic (i.e., $kT$ is significantly less than $mc^2$). As
a result, all relevant physical quantities can be described by simple
analytic expressions.

We consider throughout an isotropic radiation field. In this paragraph
we also assume a relativistic Maxwellian distribution of electrons
(which can be maintained, e.g., by Coulomb collisions). This last
assumption is dropped in \S 4.

The rate at which energy is transfered by induced Compton scattering
from an isotropic radiation field to a single electron moving at a
speed $v=\beta c$ is given by (Illarionov, Kompaneets, 1976)
\begin{eqnarray}
W^{+}(\beta) &=& \frac{12\pi\sigma_{\rm T}h^2}{mc^4}\int_0^\beta
\Phi (y^\prime)G(\beta,\beta^\prime)d\beta^\prime,
\nonumber\\
\Phi (y) &=& \int_0^\infty n(\nu)n(y\nu)\nu^4 d\nu,
\nonumber\\
G(\beta,\beta^\prime) &=& \frac{\beta^{\prime 2}}{\gamma^5\beta^8
(1+\beta^\prime)^5}\left[(30-24\beta^2+2\beta^4)\ln{\frac{y^\prime}{y}}
\right.
\nonumber\\
&&\left.
+28\beta^3-60\beta+5(3-\beta^2)^2\beta^\prime
+(5-\beta^2)(3+\beta^{\prime 2})\beta^\prime
\left(\frac{\gamma^\prime}{\gamma}\right)^2\right],
\nonumber\\
y &=&
\frac{1-\beta}{1+\beta},\,\,\,y^\prime=\frac{1-\beta^\prime}{1+\beta^\prime},
\nonumber\\
\gamma &=& 
(1-\beta^2)^{-1/2},\,\,\,\gamma^\prime=(1-\beta^{\prime 2})^{-1/2}.
\label{heat_illar}
\end{eqnarray}

Eq.~(\ref{heat_illar}) is valid for arbitrary electron
velocities, provided that $h\nu\ll mc^2$. This formula can be
simplified in some limits, in particular, for spectrally narrow
radiation lines ($\Delta\nu\ll\nu$), such as those produced in cosmic
masers and laboratory lasers (Vinogradov, Pustovalov, 1972; Blandford,
Scharlemann, 1975). We are instead interested in the
$\Delta\nu\gtrsim\nu$ case.
 
In order to calculate the instantaneous heating rate of an actual 
plasma, it is necessary to weight the result of Eq.~(\ref{heat_illar})
with a given distribution of electron velocities, $f_\beta $. The
computation procedure to determine the heating rate will therefore be 
three-dimentional integration: 1) over the radiation spectrum [when 
calculating $\Phi(y^\prime)$], 2) over $d\beta^\prime$ and 3) over
$f_\beta$. However, if the electrons obey a relativistic Maxwellian
distribution and are only mildly relativistic ($kT\lesssim 0.1 mc^2$),
it is possible to considerably simplify the computation procedure, by
reducing it to integration over the radiation spectrum, as demonstrated below.

When $kT\lesssim 0.1mc^2$, the majority of electrons have $\beta\lesssim
0.5$, and therefore the typical random Doppler frequency shift caused by a
scattering is relatively small: $\delta\nu\lesssim\nu$. The following Taylor 
expansion of the $\Phi(y^\prime)$ function entering
Eq.~(\ref{heat_illar}) is then justified:
\begin{equation}
\Phi(y^\prime)=\int_0^\infty n^2\nu^4\,d\nu+(y^\prime-1)\int_0^\infty
n\frac{\partial n}{\partial\nu}\nu^5\,d\nu
+\frac{1}{2}(y^\prime-1)^2
\int_0^\infty n\frac{\partial^2 n}{\partial\nu^2}\nu^6\,d\nu+...,
\end{equation}
where, in turn,
\begin{equation}
y^\prime-1=-2\beta^\prime+2\beta^{\prime 2}+...,\,\,\,
\frac{1}{2}(y^\prime-1)^2=2\beta^{\prime 2}+...,{\rm etc.}
\end{equation}

A similar expansion is possible for the kernel $G(\beta,\beta^\prime)$
of Eq.~(\ref{heat_illar}):
\begin{equation}
G=\frac{4}{\beta}\left[\left(\frac{\beta^\prime}{\beta}\right)^3
-2\left(\frac{\beta^\prime}{\beta}\right)^5+2\left(\frac{\beta^\prime}
{\beta}\right)^7
\right]
+20\left[-\left(\frac{\beta^\prime}{\beta}\right)^4+2\left(\frac
{\beta^\prime}{\beta}\right)^6-2\left(\frac{\beta^\prime}{\beta}\right)^8
\right]+...\,.
\end{equation}

The subsequent straightforward calculation leads to
\begin{eqnarray}
W^{+}(\beta) &=& \frac{\sigma_{\rm T}c^2}{8\pi m}
\left[J_0+\left(-\frac{17}{30}J_0-\frac{7}{5}J_1\right)\beta^2
+\left(-\frac{47}{600}J_0+\frac{17}{150}J_1+\frac{11}{30}J_2\right)\beta^4
\right.
\nonumber\\
&&\left.
+\left(-\frac{899}{19600}J_0+\frac{2833}{29400}J_1+\frac{4261}{14700}J_2
-\frac{64}{1575}J_3\right)\beta^6+...\right],
\nonumber\\
J_0 &=& \int_0^\infty\epsilon^2_\nu\nu^{-2} d\nu,\,\,
J_1=\int_0^\infty\left(\frac{\partial\epsilon_\nu}{\partial\nu}\right)^2
d\nu,
\nonumber\\
J_2 &=&\int_0^\infty\left(\frac{\partial^2\epsilon_\nu}{\partial\nu^2}\right)^2
\nu^2 d\nu
,\,\,
J_3=\int_0^\infty\left(\frac{\partial^3\epsilon_\nu}{\partial\nu^3}\right)^2
\nu^{4} d\nu.
\label{heat_beta_e}
\end{eqnarray}
Note that the Planck constant, $h$, is absent from
Eq.~(\ref{heat_beta_e}), which reflects the well-known fact that the
process of induced Compton scattering is classical (see, e.g.,
Zel'dovich, 1975).

Our next step is to replace $\beta^2$, $\beta^4$, $\beta^6$ in
Eq.~(\ref{heat_beta_e}) by their average values $\langle\beta^2\rangle$,
$\langle\beta^4\rangle$, $\langle\beta^6\rangle$, calculated for a
mildly-relativistic Maxwellian distribution at temperature $T$,
which is given by (e.g., Sazonov, Sunyaev, 2000)
\begin{eqnarray}
f_\beta=\left(\frac{2\pi kT}{mc^2}\right)^{-3/2}
\left[1+\frac{15}{8}\frac{kT}{mc^2}
+\frac{105}{128}\left(\frac{kT}{mc^2}\right)^2
-\frac{315}{1024}\left(\frac{kT}{mc^2}\right)^3+...\right]^{-1} 
\nonumber\\
\left(1+\frac{5}{2}\beta^2-\frac{3}{8}\frac{\beta^4 mc^2}{kT}
+\frac{35}{8}\beta^4-\frac{5}{4}\frac{\beta^6 mc^2}{kT}
+\frac{9}{128}\frac{\beta^8 m^2c^4}{k^2T^2}+\frac{105}{16}\beta^6
\right.
\nonumber\\
\left.
-\frac{345}{128}\frac{\beta^8 mc^2}{kT}
+\frac{75}{256}\frac{\beta^{10}m^2c^4}{k^2T^2}-\frac{9}{1024}
\frac{\beta^{12}m^3c^6}{k^3T^3}+...\right)
\exp{\left(-\frac{\beta^2 mc^2}{2kT}\right)}.
\label{fm}
\end{eqnarray}

The resulting heating rate as a function of the plasma temperature is
\begin{eqnarray}
W^{+}(T)=\frac{\sigma_{\rm T}c^2}{8\pi m}\left[J_0-\left(\frac{17}{10}J_0
+\frac{21}{5}J_1\right)\frac{kT}{mc^2}
+\left(\frac{123}{40}J_0+\frac{61}{5}J_1
+\frac{11}{2}J_2\right)\left(\frac{kT}{mc^2}\right)^2
\right.
\nonumber\\
\left.
-\left(\frac{1989}{280}J_0+\frac{453}{14}J_1+\frac{1899}{140}J_2
+\frac{64}{15}J_3\right)\left(\frac{kT}{mc^2}\right)^3+...\right],
\label{heat_temp}
\end{eqnarray}
where $J_0$, $J_1$, $J_2$ and $J_3$ were introduced in Eq.~(\ref{heat_beta_e}).

In the case of cold electrons ($kT\ll mc^2$), only the leading term of
the series in powers of $kT/mc^2$ in Eq.~(\ref{heat_temp}) is
important, and the heating rate due to induced Compton scattering
is described by the well-known (Zel'dovich, Levich, 1970; Levich,
Sunyaev, 1971) formula
\begin{equation}
W^{+}_0=\frac{\sigma_{\rm T}c^2}{8\pi m}
\int_0^\infty\epsilon^2_\nu\nu^{-2}\,d\nu.
\label{heat_temp_0}
\end{equation}

Both the nonrelativistic expression (\ref{heat_temp_0}) and the
first-order relativistic correction to it (see Eq.~[\ref{heat_temp}])
were already obtained in \S 2 following a different approach; see
Eq.~(\ref{transfer}). Thus, two independent methods give the same
result. 

%%%%%%%%%%%%%%%%%%%%%%%%%%%%%%%%%%%%%%%%%%%%%%%%%%%%%%%%%%%%%%%%%%%%%%%%
\subsection{Heating in a field of self-absorbed synchrotron radiation}
%%%%%%%%%%%%%%%%%%%%%%%%%%%%%%%%%%%%%%%%%%%%%%%%%%%%%%%%%%%%%%%%%%%%%%%%

Let us now consider one particular spectral distribution, namely
the spectrum of synchrotron radio emission with self-absorption at low
frequencies. The radiation spectrum generated by a
spherically-symmetric homogeneous source is
(Gould, 1979)
\begin{eqnarray}
\epsilon_\nu^0(\nu) &=& A\left(\frac{\nu}{\nu_0}\right)^{5/2}\left[
\frac{1}{2}+\frac{\exp{(-t)}}{t}-\frac{1-\exp{(-t)}}{t^2}\right],
\nonumber\\
t &=& \left(\frac{\nu_0}{\nu}\right)^{0.5p+2}.
\label{sync}
\end{eqnarray}
The shape of the distribution (\ref{sync}) is defined by a single
parameter, $p$, which is the index of the power-law energy
distribution of the electrons producing the synchrotron radiation,
$dN_e\sim\gamma^{-p}\,d\gamma$. The other two parameters that appear in 
Eq.~(\ref{sync}), $\nu_0$ and $A$, determine the position of
the spectrum along the frequency axis and its amplitude,
respectively. Far enough from the peak of intensity ($\nu_{\rm peak}\approx
1.15\nu_0$), the spectrum (\ref{sync}) assumes a power-law shape: 
$\epsilon_\nu^0\sim\nu^{5/2}$ in the region of self-absorption
($\nu\ll\nu_0$) and $\epsilon_\nu^0\sim\nu^{(1-p)/2}$ in the optically
thin part ($\nu\gg\nu_0$).

Since the integral $\epsilon=\int_0^\infty\epsilon_\nu^0 d\nu$ diverges
at $\infty$ for $p\le 3$, and it is this quantity that determines the
Compton cooling rate, which will be taken into account below, we
modify Eq.~(\ref{sync}) as follows:
\begin{equation}
\epsilon_\nu(\nu)=\left\{\begin{array}{ll}
  \epsilon_\nu^0(\nu), & \nu\le\nu_b\\
  (\nu/\nu_b)^{-0.5}\epsilon_\nu^0(\nu), & \nu > \nu_b.
\end{array}\right.\
\label{sync_break}
\end{equation}
Here we have assumed that a steepening of the spectrum
(increase in the slope by $0.5$) takes place above some ``break''
frequency, $\nu_b\gg\nu_0$, due to the fast synchrotron cooling of
more energetic electrons.

To be concrete, we take the slope of the optically thin part of
the spectrum (prior to the slope break) to be $\alpha=-0.7$, which
corresponds to $p=2.4$. We additionally adopt $\nu_b=10^3\nu_0$ as
a fiducial value for our treatment of electron cooling,
although this parameter just enters as a multiplicative factor in
the relevant formulae as long as $\nu_b\gg\nu_0$. The resulting
spectrum is plotted in Fig.~\ref{fig_sync}. Note that the
low-frequency breaks observed in the spectra of some radio
sources are sometimes interpreted as being caused by mechanisms
other then synchrotron self-absorption, which include free-free
absorption in the ambient medium (e.g., Bicknell et al., 1997) and
induced Compton scattering either outside or inside the radio source
(e.g., Sunyaev, 1971; Sincell, Krolik, 1994; Kuncic et al., 1998). However, the
radiation field inside the source, which we are concerned with, may 
well be the superposition of self-absorbed synchrotron spectra
generated by individual plasma components. 

Fig.~\ref{fig_rates}a shows the heating rate of thermal electrons
exposed to self-absorbed synchrotron radiation as a function of the
electron temperature. The exact result was obtained by weighting the
heating rate for monoenergetic electrons, given by
Eq.~(\ref{heat_illar}), with the relativistic Maxwellian distribution,
$dN_{\rm e}=const\,\,\,\gamma(\gamma^2-1)^{1/2}\exp{(-\gamma
mc^2/kT)}d\gamma$. We have additionally verified this dependence with
Monte-Carlo simulations (using the Comptonization code described in 
Pozdnyakov et al., 1983 with a slight modification to allow for induced
Compton scattering). The computation proves to be faster with
the semi-analytic formula (\ref{heat_illar}) of Illarionov and Kompaneets. Also
presented in Fig.~\ref{fig_rates}a are various approximations for the
heating rate that result from retaining a different number of
temperature terms in Eq.~(\ref{heat_temp}).

One can see that the heating rate begins to decrease appreciably ($>
5$\%) at $kT\sim 5$~keV. The temperature domain where
this decrement is described very well by the approximate relation
(\ref{heat_temp}) extends to $kT\sim 30$~keV, when the heating rate is
already smaller by $\sim 30$\% than in the case of cold electrons. The
further decrease in the heating rate that takes place at yet higher
temperatures cannot be described properly in the framework of the
Fokker-Planck approach, which led to Eq.~(\ref{heat_temp}). We found
it convenient to describe the exact dependence presented in
Fig.~\ref{fig_rates}a by an approximate formula, which is accurate to
within 3\% for $kT< 5mc^2$:
\begin{equation}
W^{+}(T)=\left\{0.8\exp{\left[-3.7\left(\frac{kT}{mc^2}\right)^{0.8}
\right]}
+0.2\exp{\left[-1.8\left(\frac{kT}{mc^2}\right)^{0.6}\right]}\right\}W^{+}_0,
\label{heat_fit}
\end{equation}
where, as follows from Eq.~(\ref{heat_temp_0}), (\ref{sync}:
\begin{equation}
W^{+}_0=1.1\cdot 10^{24}A^2\nu_0^{-1}\,{\rm eV\, s^{-1}}
\label{heat_temp_0_A}
\end{equation}
($A$ is measured in erg Hz$^{-1}$ cm$^{-3}$ and $\nu_0$ in GHz).

It is worth mentioning that the heating rate approaches an asymptote
$W^{+}(T)\sim (mc^2/kT)^3$ when $kT\gg mc^2$. This results from the fact that
only the lower-energy part of the relativistic Maxwellian
distribution, i.e., electrons with $\gamma\lesssim 1$, significantly
contributes to the net heating rate (because $W^{+}\sim\gamma^{-5}$ for
$\gamma\gg 1$ --- see Illarionov, Kompaneets, 1976), and the relative
number of such electrons is proportional to $(mc^2/kT)^3$.

Although we have assumed a particular slope ($-0.7$) for the
optically-thin part of the radiation spectrum, it turns out that the dependence
of the heating rate on the electron temperature changes very slowly as
the slope varies. Quantitatively, $W^{+}(T)$ remains the same to
within 10\% for $\alpha$ in the range $[-0.9; -0.5]$. Thus,
formula~(\ref{heat_fit}) is quite useful in that it allows obtaining
reasonably good estimates for the heating rate of substantially 
relativistic Maxwellian electrons in an isotropic field of
self-absorbed synchrotron  radiation.

Of course, the mildly-relativistic Eq.~(\ref{heat_temp}), which
is applicable when $kT\lesssim 30$~keV, explicitly depends on the spectral
distribution, and so can be used to calculate the heating rate
for an arbitrary (broad) radiation spectrum.

%%%%%%%%%%%%%%%%%%%%%%%%%%%%%%%%%%%%%%%%%%%%%%%%%%%%%%%%%%%%%%%%%%%%%%%%
\subsection{Stationary temperature of electrons}
%%%%%%%%%%%%%%%%%%%%%%%%%%%%%%%%%%%%%%%%%%%%%%%%%%%%%%%%%%%%%%%%%%%%%%%%

Zel'dovich and Levich (1970) studied, in the nonrelativistic
approximation, the problem about the establishment of a stationary
distribution of electrons during their interaction with an
isotropic field of low-frequency radiation of high brightness
temperature. These authors showed that if the induced Compton heating
is counteracted by cooling due to inverse (spontaneous) Compton
scattering, then the stationary distribution will be Maxwellian with a
temperature
\begin{equation}
kT_{\rm eq}^0=\frac{c^3}{32\pi\epsilon}\int_0^\infty \epsilon_\nu^2
\nu^{-2}\,d\nu,
\label{teq}
\end{equation}
where $\epsilon=\int_0^\infty \epsilon_\nu\,d\nu$.

It is reasonable to suggest that in the mildly-relativistic regime there will
be only small deviations from a Maxwellian distribution. Furthermore, if other
mechanisms, e.g., Coulomb collisions should play a significant role in
the redistribution of electrons in momentum space, then a Maxwellian
distribution can be maintained easily. We postpone a detailed
discussion of questions related to the electron distribution until \S
4. Here we will continue to assume as before that the electrons obey a
relativistic Maxwellian distribution, and will find the stationary
electron temperature, considering inverse Compton scattering (of the same
low-frequency synchrotron radiation that is heating the electrons) the only 
cooling agent. Other possible cooling mechanisms, which may prove more
important under certain conditions, will be mentioned in \S 3.3.  

The inverse Compton cooling rate is given by (e.g., Pozdnyakov et al., 1983)
\begin{equation}
W^{-}(T)=\left(\langle\gamma\rangle+\frac{kT}{mc^2}\right)W^{-}_0(T),
\label{cool_temp_gen}
\end{equation}
where the mean electron energy (in units of $mc^2$) is 
\begin{equation}
\langle\gamma\rangle =\frac{3kT K_2(mc^2/kT)+mc^2 K_1(mc^2/kT)}{2kT
K_1(mc^2/kT)+mc^2 K_0(mc^2/kT)},
\label{mean_gamma}
\end{equation}
$K_p(x)$ are modified Bessel functions, and the cooling rate in the
nonrelativistic limit ($kT\ll mc^2$) is
\begin{equation}
W^{-}_0(T)=\frac{4\sigma_{\rm T}\epsilon kT}{mc}.
\label{cool_temp_0}
\end{equation}

We can expand Eq.~(\ref{cool_temp_gen}) in powers of $kT/mc^2$ to
obtain a formula applicable in the mildly-relativistic limit, which is similar
in structure to the corresponding relation for the heating rate
(Eq.~[\ref{heat_temp}]):
\begin{equation}
W^{-}(T)=\frac{4\sigma_{\rm T}\epsilon kT}{mc^2}\left[1+\frac{5 kT}{2mc^2}+
\frac{15}{8}\left(\frac{kT}{mc^2}\right)^2
-\frac{15}{8}\left(\frac{kT}{mc^2}
\right)^3+...\right].
\label{cool_temp}
\end{equation}

An excellent fit to the exact formula (\ref{cool_temp_gen}) in the range
$kT\lesssim 5mc^2$ is provided by
\begin{equation}
W^{-}(T) = \left[1+3.4\left(\frac{kT}{mc^2}\right)^{1.1}\right]W^{-}_0(T),
\label{cool_fit}
\end{equation} 
with $W^{-}_0$ given by Eq.~(\ref{cool_temp_0}). 

The dependences $W^{-}(T)$ described by Eqs.~(\ref{cool_temp_gen})
and (\ref{cool_temp}) are plotted in Fig.~\ref{fig_rates}b and can
be compared to the corresponding dependences for the heating rate
(Fig.~\ref{fig_rates}a). We note that the relationship
$W^{-}(T)$, of course, does not depend on the shape of the radiation 
spectral distribution (depending only on the total energy density), as
opposed to $W^{+}(T)$.

We can now find the equilibrium electron temperature by solving the
equation $W^{+}(T_{\rm eq})=W^{-}(T_{\rm eq})$. In order to proceed, we need to
specify the constant $A$ appearing in Eq.~(\ref{sync}), which defines
the amplitude of the spectral distribution. It is natural to express this
coefficient through the maximal brightness temperature, $T_b^{\rm max}=
max[\epsilon_\nu(\nu)c^3/(8\pi\nu^2)]$. $T_b$ peaks at 
$\nu^{\rm max}=0.61\nu_0$ for $\alpha=-0.7$ (although the position of the
peak is only marginally dependent on $\alpha$ for typically observed
radiation spectra of extragalactic radiosources). We find:
\begin{equation}
A\approx 22\pi kT_b^{\rm max}\nu_0^2/c^3.
\label{tb}
\end{equation}

Fig.~\ref{fig_teq} shows the equilibrium electron temperature,
$T_{\rm eq}$, as a function of $T_b^{\rm max}$. The exact result is
compared with the nonrelativistic result (Eq. [\ref{teq}]) and
different-order mildly-relativistic estimates that were obtained
by equating Eq.~(\ref{heat_temp}) and Eq.~(\ref{cool_temp}). One can see
that the nonrelativistic Eq.~(\ref{teq}) is still valid for $T_{\rm
eq}\lesssim 5$~keV, which corresponds in our case to brightess
temperatures $T_b^{\rm max}\lesssim 3\cdot 10^{10}$~K. In this regime
\begin{equation}
kT_{\rm eq}^0=1.9\frac{T_b^{\rm max}}{10^{10}{\rm K}}
\left(\frac{\nu_b}{1000\nu_0}\right)^{-0.33}\,{\rm keV}.
\label{teq_sync}
\end{equation}
The next decade of values of the equilibrium temperature, up to
$kT_{\rm eq}\sim 40$~keV, is well described by the approximate formulae
(\ref{heat_temp}) and (\ref{cool_temp}). Note that $kT_{\rm eq}=40$~keV 
corresponds to $T_b^{\rm max}\sim 4\cdot 10^{11}(\nu_b/1000\nu_0)^{0.33}$~K.
An important conclusion can be made: {\sl electrons can
be heated up to mildly relativistic temperatures, $kT\sim$ a few tens
of keV, as a result of induced Compton scattering of synchrotron
radiation with $T_b\sim 10^{11}$--$10^{12}$~K, but not above these
temperatures}.

%%%%%%%%%%%%%%%%%%%%%%%%%%%%%%%%%%%%%%%%%%%%%%%%%%%%%%%%%%%%%%%%%%%%%%%%%%%%%
\subsection{Evolution of the electron temperature during Compton interaction}
%%%%%%%%%%%%%%%%%%%%%%%%%%%%%%%%%%%%%%%%%%%%%%%%%%%%%%%%%%%%%%%%%%%%%%%%%%%%%

Let us now address a related timing problem, namely, examine how
rapidly electrons can be heated to mildly relativistic temperatures
through induced Compton scattering.

We will first estimate the basic characteristic quantities in the
nonrelativistic limit. Substituting $A$ given by Eq.~(\ref{tb}) into
Eq.~(\ref{heat_temp_0_A}), we find
\begin{equation}
W^{+}_0= 1.4\cdot 10^{-9}\left(\frac{T_b^{\rm
max}}{10^{11}{\rm K}}\right)^2\left(\frac{\nu_0}{1\rm{GHz}}\right)^3
{\rm eV\, s^{-1}}.
\label{heat_abs_0}
\end{equation}

If there were no cooling, then initially cold electrons would acquire
a kinetic energy of $kT_{\rm eq}^0$, given by Eq.~(\ref{teq_sync}),
during
\begin{equation}
t_{\rm heat}=\frac{kT_{\rm eq}^0}{W^{+}_0}= 1.4\cdot 10^{13}
\left(\frac{T_b^{\rm max}}{10^{11}{\rm K}}\right)^{-1}
\left(\frac{\nu_0}{1{\rm GHz}}\right)^{-3}
\left(\frac{\nu_b}{1000\nu_0}\right)^{-0.33} {\rm s}.
\label{time_heat}
\end{equation}
We notice that the heating time depends strongly on the characteristic
frequency of the synchrotron self-absorption: $t_{\rm heat}\sim\nu_0^{-3}$. 

Using Eq.~(\ref{cool_temp_0}), we can find the corresponding cooling
rate:
\begin{equation}
W^{-}_0(T)=3.8\cdot 10^{-8}\frac{kT}{mc^2}\frac{T_b^{\rm max}}
{10^{11}{\rm K}}\left(\frac{\nu_0}{1\rm{GHz}}\right)^3
\left(\frac{\nu_b}{1000\nu_0}\right)^{0.33}{\rm eV\, s^{-1}}.
\label{cool_abs_0}
\end{equation}

In the mildly-relativistic regime, the heating and cooling
rates will be respectively smaller and larger than those
given by Eqs.~(\ref{heat_abs_0}) and (\ref{cool_abs_0}). We have
computed the evolution of the electron temperature for a set of
$T_b^{\rm max}$ values by integrating the following differential equation:
\begin{equation}
\frac{dT}{dt}=\left(\frac{d\langle\gamma\rangle(T)}{dT}\right)^{-1}
\frac{W^{+}(T)-W^{-}(T)}{mc^2},
\end{equation}
using Eq.~(\ref{mean_gamma}) to represent the dependence
$\langle\gamma\rangle(T)$ and the fitting formulae (\ref{heat_fit}) and
(\ref{cool_fit}) for $W^{+}(T)$ and $W^{-}(T)$, respectively.
The resulting time histories are presented in
Fig.~\ref{fig_temp_time}, for which the value $\nu_0=1$~GHz was
used. For a given value of $\nu_0$, one should simply rescale the time
axis in Fig.~\ref{fig_temp_time} as $(\nu_0/1{\rm GHz})^{-3}$.

Each of the time histories presented in Fig.~\ref{fig_temp_time} can be
divided into two intervals. During the earlier period, the temperature
grows linearly because $W^{+}\gg W^{-}$. As $T$ becomes $\gtrsim 0.5 T_{\rm
eq}$, the second (longer) period starts, during which the cooling
plays a major role and the temperature slowly approaches the equilibrium
value. This transition period is additionally lengthened by
relativistic effects (compare the different solutions in
Fig.~\ref{fig_temp_time}). We can define a characteristic time of 
induced heating as the time needed to heat the plasma to $\sim
0.5T_{\rm eq}$. To give an example, for $T_b^{\rm max}= 
10^{12}$~K, a time of $\sim 3\cdot 10^{4}(\nu_0/1\rm{GHz})^{-3}$~years is
required to heat the electrons  to $kT=0.5kT_{\rm eq}\sim 34$~keV. If further,
$\nu_0\sim 10\rm{GHz}$, the heating time becomes $\sim
30$~years. Interestingly enough, the simple nonrelativistic
Eq.~(\ref{time_heat}) provides a good estimate (within a factor of 2)
for the heating time even for values of $T_b^{\rm max}$ as high as $\sim
10^{13}(\nu_b/1000\nu_0)^{0.33}$~K (of course, the equilibrium
temperature in this case is much less than the nonrelativistic estimate).

In real situations, there may be mechanisms operating by which the
plasma cools more efficiently than by inverse Compton scattering of
the synchrotron radiation. One should then modify accordingly the
cooling rate $W^{-}(T)$ in our treatment above. If the energy density
of a possible high-frequency radiation component is larger than that of the
low-frequency synchrotron emission, then the contribution of this component
to the inverse Compton cooling rate will be accordingly larger. 
Also, cooling due to free-free transitions will become more 
important than inverse Compton cooling if the plasma is 
dense enough, namely when $N_{\rm e}T^{-1/2}\epsilon^{-1} > 10^4$
K$^{-1/2}$ erg$^{-1}$. For the synchrotron spectrum described by
Eqs.~(\ref{sync}) and ({\ref{sync_break}), this condition translates
into  $N_{\rm e}> 6(T_b^{\rm max}/10^{11}{\rm K})^{3/2}(\nu_0/1{\rm GHz})^{3}
(\nu_b/1000\nu_0)^{0.166}$~cm$^{-3}$.
Another possible cooling mechanism is adiabatic expansion of a
plasma cloud. The characteristic time scale for this 
process is $t_{\rm ad}=3\cdot 10^{10}(R/1{\rm pc})(U/10^3
{\rm km\,s}^{-1})^{-1} s$, where $R$ is the size of the cloud and $U$
is the expansion velocity (assuming spherical expansion). 

%%%%%%%%%%%%%%%%%%%%%%%%%%%%%%%%%%%%%%%%%%%%%%%%%%%%%%%%%%%%%%%%%%%%%%%%
\section{Effect of induced Compton scattering on the electron distribution} 
%%%%%%%%%%%%%%%%%%%%%%%%%%%%%%%%%%%%%%%%%%%%%%%%%%%%%%%%%%%%%%%%%%%%%%%%

So far we have assumed that the distribution of the electrons in
momentum space remains relativistic Maxwellian during their interaction
with the low-frequency radiation. This should be the
case if the plasma is dense enough that electron-electron
collisions can quickly smooth out any arising deviations from a
Maxwellian distribution. In order to find out whether this
thermalization really takes place, one should compare the induced
Compton heating time, $t_{\rm heat}$, given by Eq.~(\ref{time_heat}), with the
time scale on which electrons can relax to a Maxwellian distribution
(Spitzer, 1978), $t_{\rm e-e}=2.5\cdot
10^{12}(\ln{\Lambda}/40)^{-1}(kT/mc^2)^{3/2}N_{\rm e}^{-1} $~s, where 
$\ln{\Lambda}$ is the Coulomb logarithm. Taking $T=T_{\rm eq}^0$
(Eq.~[\ref{teq_sync}]), we find that relaxation is efficient when
\begin{equation}
\frac{t_{\rm e-e}}{t_{\rm heat}}=10^{-3}
\left(\frac{\ln{\Lambda}}{40}\right)^{-1}
\left(\frac{T_b^{\rm max}}{10^{11}{\rm K}}\right)^{5/2}
\left(\frac{\nu_0}{1{\rm GHz}}\right)^{3}
\left(\frac{\nu_b}{1000\nu_0}
\right)^{-0.166}N_{\rm e}^{-1}<1.
\end{equation}

Let us consider a few examples. At $T_b^{\rm max}=10^{11}$~K and 
$\nu_0=1$~GHz, Maxwellization of the electron spectrum occurs when
$N_{\rm e}\gtrsim 10^{-3}$~cm$^{-3}$. For $T_b^{\rm max}=5\cdot 10^{11}$~K and
$\nu_0=10$~GHz, the corresponding range is $N_{\rm e}\gtrsim 10^2$~cm$^{-3}$. 
It should be noted here that the establishment of the high-velocity tail of
the Maxwellian distribution occurs on a larger time scale than
$t_{\rm e-e}$ given above. In the nonrelativistic limit, the appropriate
characteristic time is proportional to $(v/\langle v\rangle)^3$ when
$v\gg\langle v\rangle$, where $v$ is the velocity of an electron 
and $\langle v\rangle$ is the typical thermal velocity of electrons
(e.g., Krall, Trivelpiece, 1973).

We now consider an extreme situation when $t_{\rm e-e}\gg
t_{\rm heat}$, i.e., no Maxwellization of electrons takes place due to
collisions. To find out what momentum distribution, $f(p)$, results in
this case, we need to consider the diffusion of electrons in momentum space
caused by induced and spontaneous (inverse) Compton scattering. The 
corresponding Fokker-Planck equation for the electron momentum
distribution is (Illarionov, Kompaneets, 1976) 
\begin{equation}
\frac{\partial f}{\partial t}=\frac{1}{p^2}\frac{\partial}{\partial p}
p^2\left[D\frac{\partial f}{\partial p}-F_{\rm sp}f\right],
\label{diff_illar}
\end{equation}
where the diffusion coefficient
\begin{eqnarray}
D(\beta) &=& \frac{12\pi\sigma_{\rm T}h^2}{c^4}\int_0^\beta
\Phi (y^\prime)G_D(\beta,\beta^\prime)d\beta^\prime\,
\nonumber\\
G_D(\beta,\beta^\prime) &=& \frac{2\beta^{\prime 2}}{\gamma^4\beta^8
(1+\beta^\prime)^5}\left[(3-\beta^2)\ln{\frac{y}{y^\prime}}+2\beta
-2\beta^\prime\left(\frac{\gamma^\prime}{\gamma}\right)^2
+2(\gamma^2+\gamma^{-2})(\beta-\beta^\prime)\right],
\nonumber\\
\label{d_illar}
\end{eqnarray}
and the breaking force due to spontaneous scattering (Landau \& Lifshits 1975)
\begin{equation}
F_{\rm sp}(\beta)=-\frac{4}{3}\sigma_{\rm T}\epsilon\beta\gamma^2.
\label{fsp}
\end{equation}
The electron velocity is related to the electron momentum via
\begin{equation}
\beta=\frac{p}{\gamma mc}=\frac{p}{mc}
\left[1+\left(\frac{p}{mc}\right)^2\right]^{-1/2}.
\end{equation}
The remaining quantities appearing in Eqs.~(\ref{d_illar}) and
(\ref{fsp}) are the same as in Eq.~(\ref{heat_illar}). 

The cumbersome expression (\ref{d_illar}) can be simplified
if we assume that $\beta\ll 1$. Series expansion can then be
carried out, which is completely analogous to those written down in \S
3 for the induced Compton heating rate:
\begin{equation}
D(\beta)=\frac{\sigma_{\rm T}c^2}{24\pi}
\int_0^\infty\epsilon^2_\nu\nu^{-2}d\nu\left[1+\beta^2\left(\frac{4}{25}
-\frac{21\int_0^\infty(\partial\epsilon_\nu/\partial\nu)^2\,d\nu}
{25\int_0^\infty\epsilon^2_\nu\nu^{-2}d\nu}\right)+...\right].
\label{d_beta}
\end{equation}

The equilibrium momentum distribution is given by
\begin{equation}
f_{\rm eq}(p)=const\,\,\,\exp{\left[\int_0^p
\frac{F_{\rm sp}(p^\prime)}{D(p^\prime)}dp^\prime\right]},
\label{feq_illar}
\end{equation}

In the nonrelativistic limit, $D(p)=const$ and $f_{\rm sp}(p)\sim p$;
hence $F_{\rm eq}\sim\exp(-p^2/2mkT_{\rm eq}^0)$ with $T_{\rm eq}^0$
given by Eq.~(\ref{teq}). Therefore, the equilibrium distribution is
Maxwellian in this limit, which was first shown by Zel'dovich and
Levich (1970).

We can find the first-order relativistic correction to the
nonrelativistic equilibrium distribution from Eq.~(\ref{feq_illar}),
by making use of the approximate expression (\ref{d_beta}) for the
diffusion coefficient and transforming Eq.~(\ref{fsp}) to $F_{\rm
sp}=-4/3\,\sigma_{\rm T}\epsilon(p/mc)[1+0.5(p/mc)^2+...]$. The result is 
\begin{equation}
f_{\rm eq}(p)=const\,\,\,\exp \left\{-\frac{p^2}{2mkT_{\rm eq}^0}
\left[1+\left(\frac{p}{mc}\right)^2\left(\frac{17}{100}
+\frac{21\int_0^\infty(\partial\epsilon_\nu/\partial\nu)^2 d\nu} 
{50\int_0^\infty\epsilon_\nu^2\nu^{-2}d\nu}\right)\right]\right\},
\label{feq_p2}
\end{equation}
where $T_{\rm eq}^0$ is given by Eq.~(\ref{teq}). 

In the case of synchrotron radiation with self-absorption at low
frequencies (Eqs.~[\ref{sync}], [\ref{sync_break}]), the equilibrium
distribution is \begin{equation}
f_{\rm eq}(p)=const\,\,\,\exp{\left\{-\frac{p^2}{2mkT_{\rm eq}^0}
\left[1+0.69\left(\frac{p}{mc}\right)^2\right]\right\}}.
\label{feq_fit}
\end{equation}

In Fig.~\ref{fig_feq}, we have plotted the equilibrium distributions
for two values of $T_b^{\rm max}$: $5\cdot 10^{11}$~K 
and $5\cdot 10^{12}$~K. The corresponding stationary temperatures, as
estimated using nonrelativistic Eq.~(\ref{teq}), are $kT_{\rm
eq}=95$~keV and 950~keV. The exact result was obtained by numerical
evaluation of Eq.~(\ref{feq_illar}) using Eqs.~(\ref{d_illar}) and
(\ref{fsp}). Also plotted are the nonrelativistic Maxwellian
distribution at temperature $T_{\rm eq}^0$ and the mildly-relativistic
approximation given by Eq.~(\ref{feq_fit}). One can see that the right
wing  of the distribution is substantially suppressed
compared to the nonrelativistic Maxwellian distribution. This takes
place because of the decreasing diffusion coefficient and increasing
breaking force with increasing $p$. Surprisingly enough, the
approximate formula (\ref{feq_fit}) provides an excellent fit even when
$kT_{\rm eq}^0\gtrsim mc^2$.

Since the distribution (\ref{feq_illar}) becomes a Maxwellian
one in the limit $kT_{\rm eq}^0\ll mc^2$ and assumes a
quasi-Maxwellian shape in the mildly-relativistic regime, it is
natural to characterize this type of distribution by some effective
temperature, $T_{\rm eff}$. We define $T_{\rm eff}$ to be the
temperature of the relativistic Maxwellian distribution for which the
mean electron energy, $\langle\gamma\rangle
mc^2$=$\langle(p^2+1)^{1/2}\rangle$, is equal to that for a given
quasi-Maxwellian distribution, i.e.,
\begin{equation}
\frac{\int (p^2+1)^{1/2} p^2 f_{\rm eq}(p)dp}{\int p^2
f_{\rm eq}(p)dp}=\langle\gamma\rangle (T_{\rm eff})mc^2,
\label{teff}
\end{equation}
where the dependence of $\langle\gamma\rangle$ on temperature is given by
Eq.~(\ref{mean_gamma}).

For the two example distributions presented in Fig.~\ref{fig_feq},
$kT_{\rm eff}$ takes values of 45 and 140~keV --- these should be
compared with $kT_{\rm eq}^0=95$ and 950~keV, respectively. The
relativistic Maxwellian distributions that correspond to these $T_{\rm
eff}$ values are shown in Fig.~\ref{fig_feq}.

In Fig.~\ref{fig_teff}, we have plotted $T_{\rm eff}$, calculated
using both the exact formula (\ref{feq_illar}) and its
mildly-relativistic approximation (\ref{feq_p2}), as a function of
$T_b^{\rm max}$. For comparison, the dependence $T_{\rm eq} (T_b^{\rm
max})$ for Maxwellian electrons is reproduced from
Fig.~\ref{fig_teq}. One can see that the two exact dependencies are
nearly coincident (the difference is less than 10\%) in an extremely
broad range of parameter values: $kT_{\rm eff}, kT_{\rm eq}\lesssim mc^2$. Even
more surprising is the nearly perfect agreement between the exact
solution for $T_{\rm eq} (T_b^{\rm max})$ and the midly-relativistic
approximation (Eq.~[\ref{feq_p2}]) for $T_{\rm eff} (T_b^{\rm max})$. We
should emphasize here that the differences between the different
dependences shown in Fig.~\ref{fig_teff} are real (as is confirmed by
the fact that they diverge significantly when $T_{\rm eff}\gg mc^2$), 
although very small. An important conclusion follows: the energy
that can be accumulated by an ensemble of electrons as a result of induced
Compton heating (with inverse Compton scattering serving to cool the
electrons) is almost independent in the sub-relativistic regime on
whether the electrons are maintained Maxwellian while being heated or
not.

We have also checked that the contribution of the heated electrons to
the gas pressure, which is proportional to $\langle p\beta\rangle\sim\int
(p^2+1)^{1/2}[1-(1+p^2)^{-1}] p^2 f(p)dp$ (compare with
Eq.~[\ref{teff}] for the mean energy), proves to be nearly the same
(to within 2\% for arbitrary values of $T_{\rm eff}$) for a
quasi-Maxwellian plasma with the distribution function (\ref{feq_illar})
and for the thermal plasma with $T=T_{\rm eff}$ (note that in the
limit $kT_{\rm eff}\gg mc^2$ both pressures must be equal, because
$\beta\rightarrow 1$). Therefore, the effective temperature $T_{\rm
eff}$ perfectly characterizes the thermodynamic properties of a
mildly relativistic quasi-Maxwellian plasma that has been heated by
means of the induced Compton process.

The above discussion also suggests that one can make use of the simple
expression (\ref{feq_p2}) to estimate the mean stationary electron
energy with high accuracy in a very broad parameter range, at least up
to $T_{\rm eff}\sim mc^2$, which for our model spectrum
corresponds to as high as $T_b^{\rm max}\sim 10^{15}$~K,
independent of whether thermalization takes place or not.

Suppose that plasmas whose electrons are in the distribution given
by Eq.~(\ref{feq_p2}) do exist. Would they be different from thermal
plasmas observationally? In particular, will the energy spectrum of
hard X-ray bremsstrahlung emission from such plasmas be peculiar? We
have computed such a spectrum for the momentum distribution shown in 
Fig.~\ref{fig_feq}a ($kT_{\rm eff}=45$~keV). The result is
presented in Fig.~\ref{fig_bremss}. The computation consisted of weighting the
Bethe-Heitler formula (Jauch, Rohrlich, 1976) for the cross-section
of electron-ion bremsstrahlung with the given momentum distribution
(Eq.~[\ref{feq_fit}]). One can see that the deviation from the
spectrum produced by an ensemble of relativistic Maxwellian electrons
with temperature $T_{\rm eff}$ is negligible. Only when $kT_{\rm
eff}\gtrsim 100$~keV, does the bremsstrahlung spectrum formed in the
quasi-Maxwellian plasma become noticeably different from that
corresponding to a thermal plasma.

This research was partially supported by the Russian Foundation for
Basic Research (projects 00-02-16681 and 00-15-96649). 

\section*{Appendix}

In a recent paper (Sazonov, Sunyaev, 2000) we have given an analytic
expression for the kernel of the integral kinetic equation describing
the redistribution of photons in frequency as a result of induced
Compton scattering in a mildly relativistic thermal plasma. This
kernel allows one, in principle, to derive the terms associated with
the induced Compton process in Eq.~(\ref{kgen}) --- the generalized
Kompaneets equation. Unfortunately, the published expression contains
a minor error (in the leading coefficient), and as a consequence, the
formula does not correspond to the stated accuracy (it is valid when
$kT\lesssim 0.01mc^2$, instead of $kT\lesssim 0.1mc^2$). We use the
opportunity to give here the correct expression:
\begin{eqnarray}
K^{\rm ind}(\nu,\bg{\Omega};\nu^\prime,\bg{\Omega^\prime})
 &=&  \left(\frac{\nu^\prime}{\nu}\right)^2
K(\nu^\prime,\bg{\Omega^\prime}\rightarrow\nu,\bg{\Omega})-K(\nu,\bg{\Omega}
\rightarrow\nu^\prime,\bg{\Omega^\prime})
\nonumber\\
& = & \frac{3}{32\pi}\sqrt{\frac{2}{\pi}}\left(\frac{kT}{mc^2}\right)^{-3/2}
\frac{h\nu^\prime (\nu^\prime-\nu)}{mc^2 g\nu}
\left[1+\mu^2+\left(\frac{1}{8}-\mu-\frac{63}{8}\mu^2+5\mu^3\right)
\frac{kT}{mc^2}
\right.
\nonumber\\
&&\left.
-\mu(1-\mu^2)\left(\frac{\nu^\prime-\nu}{g}\right)^2
-\frac{3(1+\mu^2)mc^2}{8kT}\left(\frac{\nu^\prime-\nu}{g}\right)^4\right]
\exp{\left[-\frac{(\nu^\prime-\nu)^2mc^2}{2g^2kT}\right]},
\nonumber\\
g &=& |\nu\bg{\Omega}-\nu^\prime\bg{\Omega^\prime}|
=(\nu^2-2\nu\nu^\prime\mu+\nu^{\prime 2})^{1/2}.
\label{k_ind}
\end{eqnarray}
Here, $\nu$ and $\bg{\Omega}$ are the frequency and direction of
propagation of a photon before scatter, $\nu^\prime$ and
$\bg{\Omega}^\prime$ are the corresponding values after scatter, and
$\mu=\bg{\Omega}\bg{\Omega^\prime}$ is the scattering angle.

An analogous correction is due for the kernel averaged over the
scattering angle (see Sazonov, Sunyaev, 2000):
\begin{eqnarray}
P^{\rm ind}(\nu;\nu^\prime) &=& \left(\frac{\nu^\prime}{\nu}\right)^2
P(\nu^\prime\rightarrow\nu)-P(\nu\rightarrow\nu^\prime)
=2\sqrt{\frac{2}{\pi}}\left(\frac{kT}{mc^2}\right)^{-3/2}
\frac{h\nu^\prime(\nu^\prime-\nu)}{mc^2\nu(\nu+\nu^\prime)}
(p_0+p_t),\
\nonumber\\
p_0 &=& \left(\frac{11}{20}+\frac{4}{5}\delta^2+\frac{2}{5}\delta^4\right)F
+|\delta|\left(-\frac{3}{2}-2\delta^2-\frac{4}{5}\delta^4\right)G,
\nonumber\\
p_t &=& \left[\left(-\frac{1091}{1120}-\frac{507}{560}\delta^2
+\frac{57}{35}\delta^4+\frac{68}{35}\delta^6\right)F+|\delta|\left(\frac{9}{4}
+\delta^2-\frac{26}{5}\delta^4-\frac{136}{35}\delta^6\right)G\right]
\frac{kT}{mc^2},
\nonumber\\
F &=& \exp{(-\delta^2)},\,\,\,G=\int_{|\delta|}^{\infty}\exp{(-t^2)}\,dt=0.5
\pi^{1/2}Erfc(|\delta|),
\nonumber\\
\delta &=& \left(2\frac{kT}{mc^2}\right)^{-1/2}
\frac{\nu^\prime-\nu}{\nu^\prime+\nu}.
\label{p_ind}
\end{eqnarray}

The formulae (\ref{k_ind}) and (\ref{p_ind}) are applicable in the
range $h\nu\ll kT\lesssim 0.1mc^2$, $nh\nu=kT_b\gg kT$ (the latter
condition means that the inverse Compton effect is small with
respect to the induced one).

\section*{References}

Bicknell G.F., Dopita M.A., O'Dea C.P.O.// Astrophys. J., 1997, 485, 112

Blandford R.D.// Astron. Astrophys., 1973, v. 26, p. 161.

Blandford R.D., Scharlemann E.T.// Astrophys. Space Science, 1975,
v. 36, p. 303.

Challinor A., Lasenby A.// Astrophys. J., 1998, v. 499, p. 1.

Celotti A., Kuncic Z., Rees M.J., Wardle J.F.C.//
Mon. Not. R. Astron. Soc., 1998, v. 293, p. 288.

Gould R.J.// Astron. Astrophys, 1979, v. 76, p. 306.

Illarionov A.F., Kompaneets D.A.// Sov. Phys. JETP, 1976, v. 44, p. 930.

Itoh N., Kohyama Y., Nozawa S.// Astrophys. J., 1998, v. 502, p. 7.

Jauch J.M., Rohrlich F.// The Theory of Photons and
Electrons 2nd ed.), New York: Springer-Verlag, 1976.

Kellerman K.I., Vermeulen R.C., Zensus J.A., Cohen M.H.//
Astrophys. J., 1998, v. 115, p. 1295. 

Kompaneets A.S.// Sov. Phys. JETP, 1957, v. 4, p. 730. 

Krall N.A., Trivelpiece A.W.// Principles of plasma
physics, New York: McGraw-Hill, 1973.

Kuncic Z., Bicknell G.V., Dopita M.A.// Astrophys. J.,
1998, v. 495, L35. 

Landau L.D., Lifshitz E.M.// The Classical Theory of Fields (4th ed.),
Oxford: Pergamon press, 1975.

Levich E.V., Sunyaev R.A.// Sov. Astron., 1971, v. 15, p. 363.

Peyraud J.// J. de Phys., 1968, v. 29, p. 88.

Pozdnyakov L.A., Sobol I.M., Sunyaev R.A.// Astrophys. Space
Phys. Rev., 1983, v. 2, p. 189.

Sazonov S.Y., Sunyaev R.A.//  Astrophys. J., 2000,
v. 543, p. 28.

Shestakov A.I., Kershaw D.S., Prasad M.K.//
J. Quant. Spectr. Rad. Transf., 1988, v. 40, p. 577.

Sincell M.W., Krolik J.H.// Astrophys. J., 1994,
v. 430, p. 550.

Spitzer L.// Physical Processes in the Interstellar Medium,
Chichester: Wiley, 1978.

Sunyaev R.A.// Sov. Astron., 1971, v. 15, p. 190. 

Vermeulen R.// in Proc. ``Astrophysical Jets'', Cambridge:
Cambridge Univ. Press (eds. Burgarella D. et al.), p. 241, 1992.

Vinogradov A.V., Pustovalov V.V.// Sov. Phys. JETP, 1972, v. 35, p. 517. 

Woodward P.// Phys. Rev. D, 1970, v. 1, p. 2731.

Zel'dovich Ya.B.// Sov. Phys. Usp., 1975, v. 18, p. 79.

Zel'dovich Ya.B., Levich E.V.// Sov. Phys. JETP Lett., 1970, v. 11,
p. 35. 

%\clearpage

\begin{figure}
\epsfxsize=17cm
\epsffile{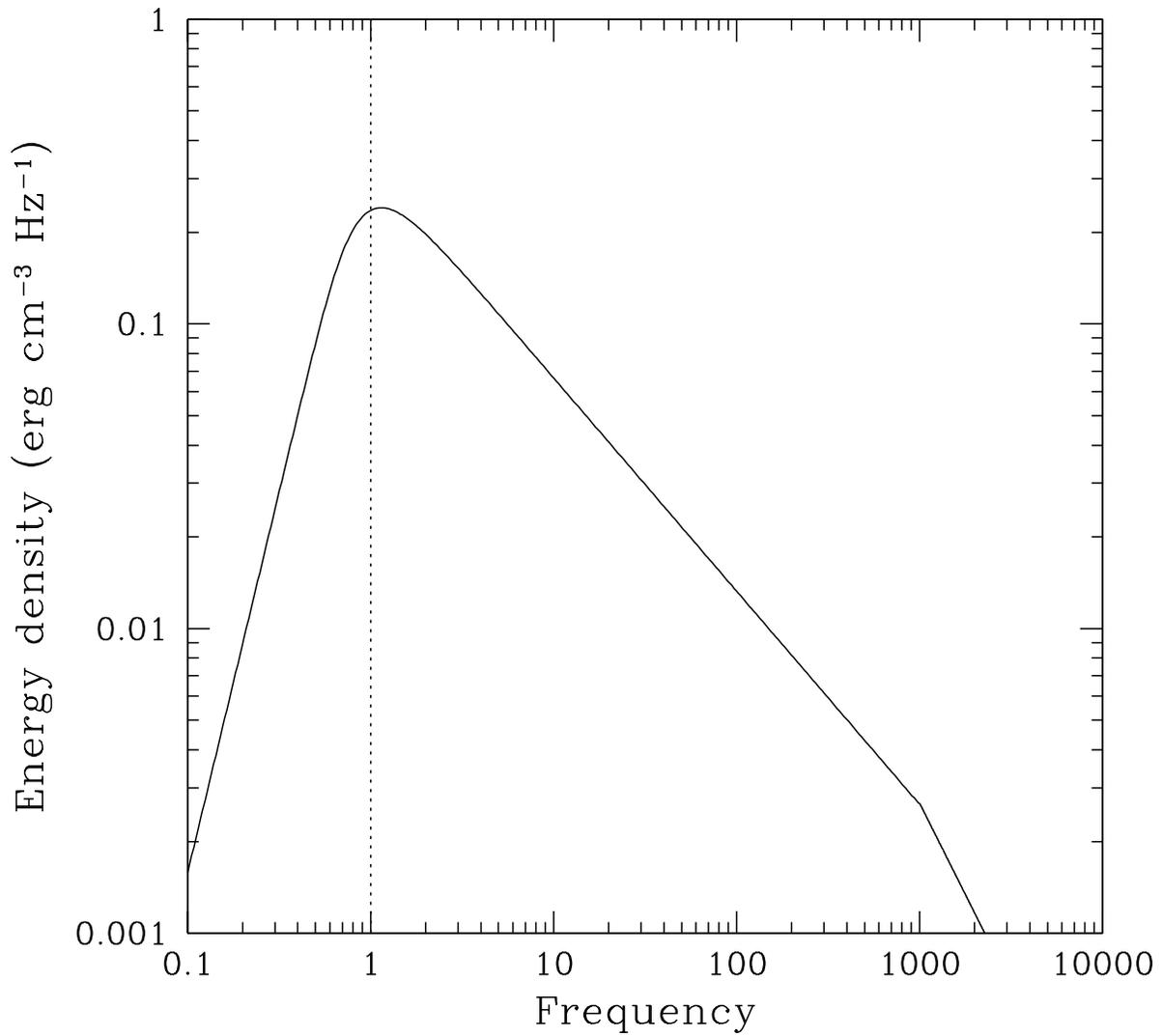}
\caption{Radiation spectrum produced by a self-absorbed synchrotron
source, described by Eqs.~(\ref{sync}) and (\ref{sync_break}). The frequency
is measured in units of the characteristic frequency $\nu_0$.
}
\label{fig_sync}
\end{figure}

\begin{figure}
\epsfxsize=17cm
\epsffile{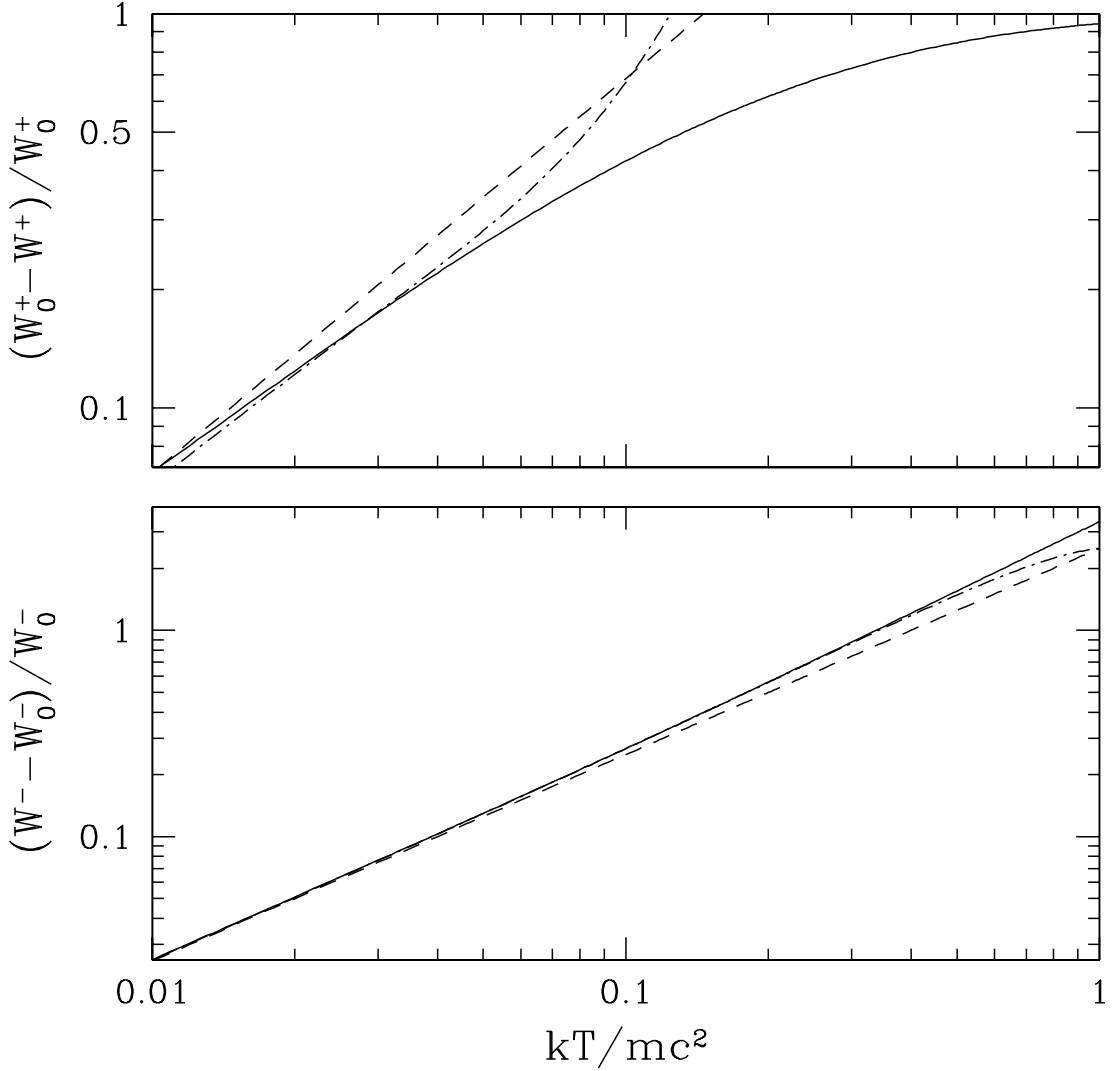}
\caption{(a) Deviation of the induced Compton heating rate of
Maxwellian electrons located in an isotropic field of self-absorbed synchrotron
radiation from the nonrelativistic estimate, $W^{+}_0$
(Eq.~[\ref{heat_temp_0}]), as a function of temperature. The solid 
line --- the exact result, obtained by weighting
Eq.~(\ref{heat_illar}) with a relativistic Maxwellian 
distribution. The dashed and dash-dotted lines represent the  
results of the calculation by the mildly-relativistic formula
(\ref{heat_temp}) in which retained are, respectively, only
the correction term $O(kT/mc^2)$ and all quoted terms up to
$O((kT/mc^2)^3)$. (b) Deviation of the inverse Compton cooling 
rate from the nonrelativistic estimate, $W^{-}_0 (T)$
(Eq.~[\ref{cool_temp_0}]). The solid line --- the exact result
(Eq.~[\ref{cool_temp_gen}]). The dashed and dash-dotted lines 
represent the results of the calculation by the mildly-relativistic
formula (\ref{cool_temp}), similarly as in (a).
}
\label{fig_rates}
\end{figure}

\begin{figure}
\epsfxsize=17cm
\epsffile{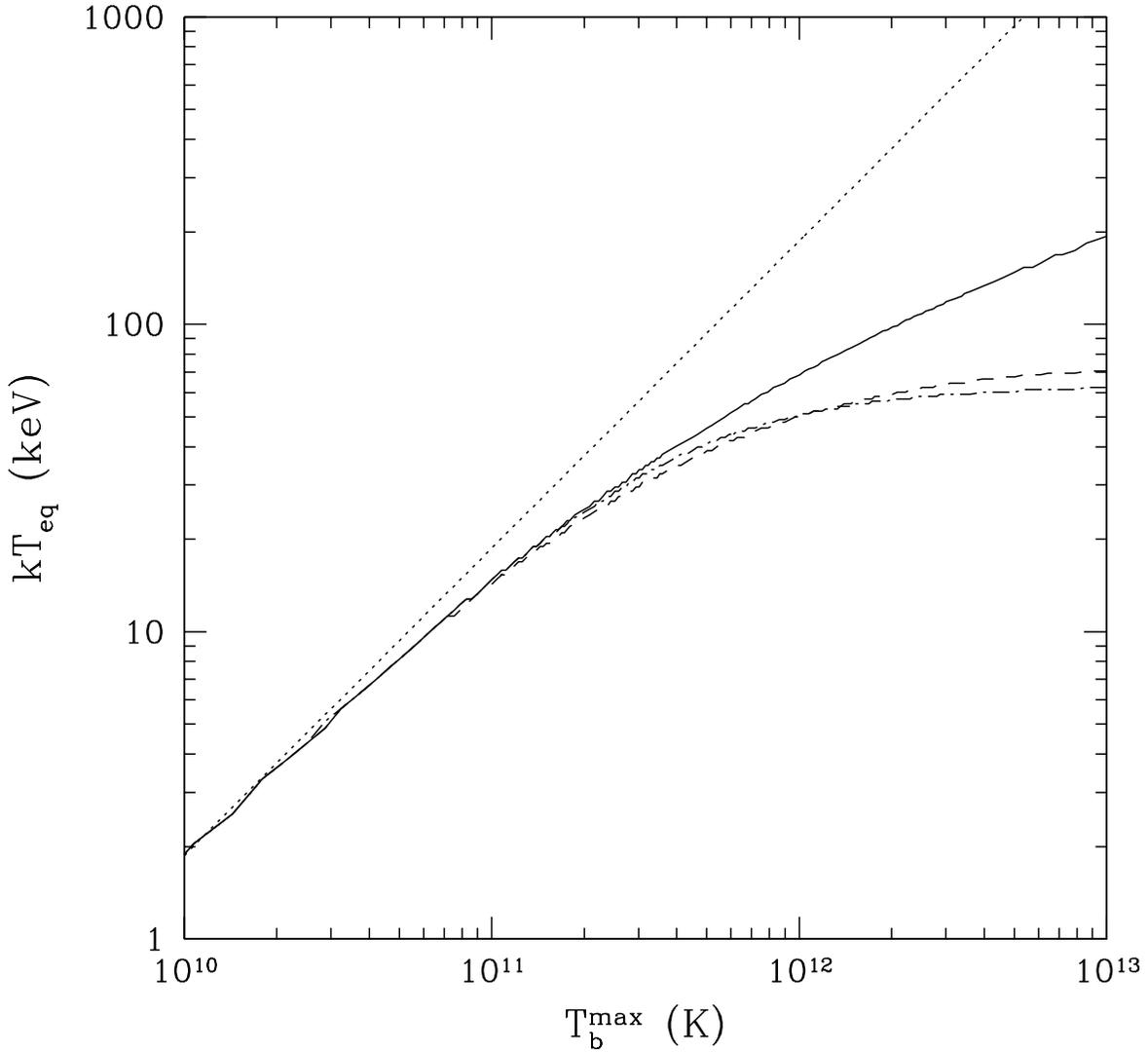}
\caption{Stationary electron temperature vs. peak radiation brightness
temperature. This dependence results from the balance $W^{+}(T_{\rm
eq})=W^{-}(T_{\rm eq})$, with $W^{+}(T)$ and $W^{-}(T)$ as plotted in
Fig.~\ref{fig_rates}. The types of the lines have the same meaning as
in Fig.~\ref{fig_rates}. Also shown (the dotted line) is the
nonrelativistic result (Eq.~[\ref{teq}] or Eq.~[\ref{teq_sync}]).
}
\label{fig_teq}
\end{figure}

\begin{figure}
\epsfxsize=17cm
\epsffile{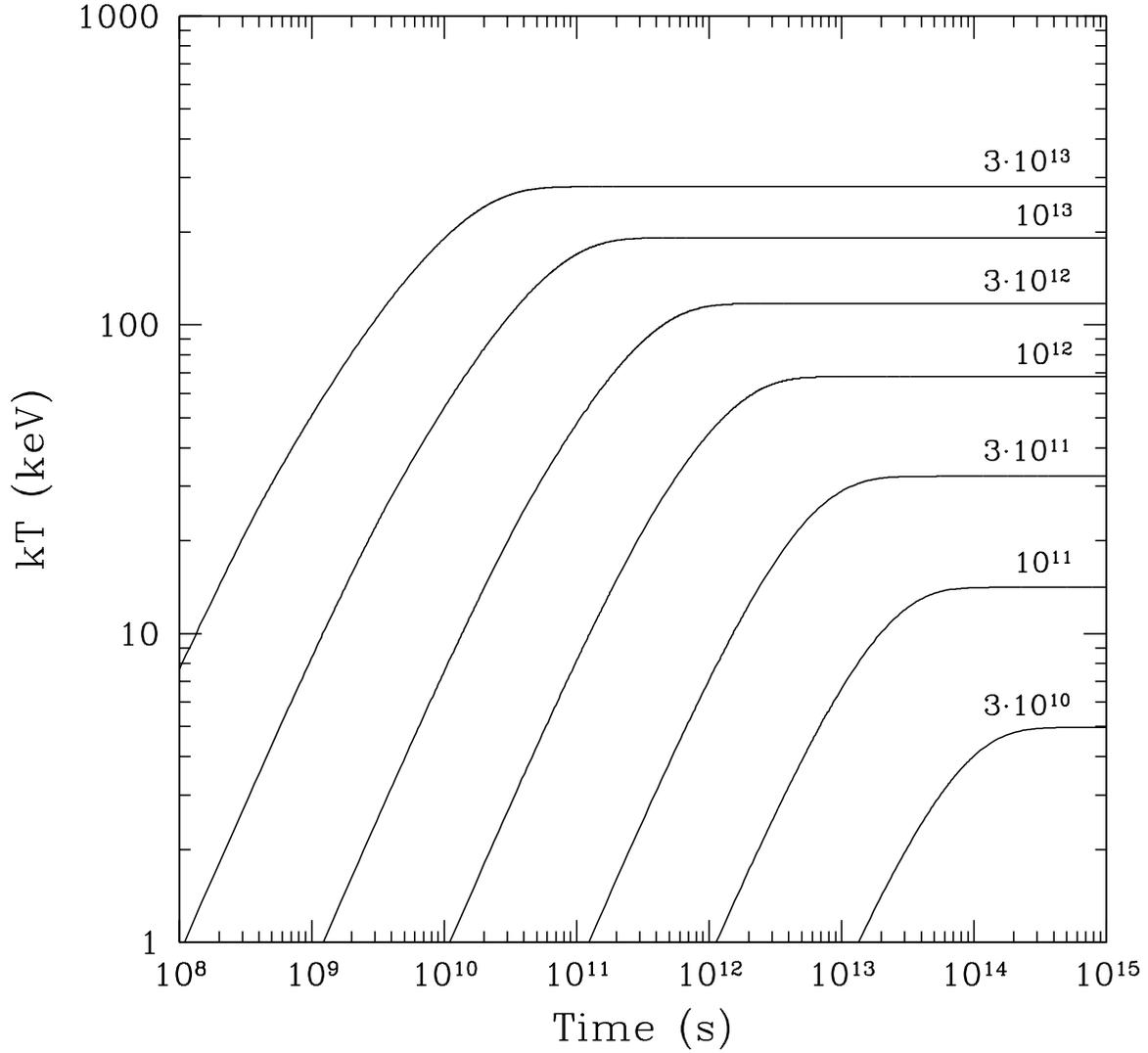}
\caption{Evolution of the temperature of Maxwellian electrons
during their Compton interaction with self-absorbed synchrotron
radiation. The curves are labelled with the corresponding values (in
K) of the peak brightness temperature $T_b^{\rm max}$. At moment $t=0$
the plasma is cold.
}
\label{fig_temp_time}
\end{figure}

\begin{figure}
\epsfxsize=17cm
\epsffile{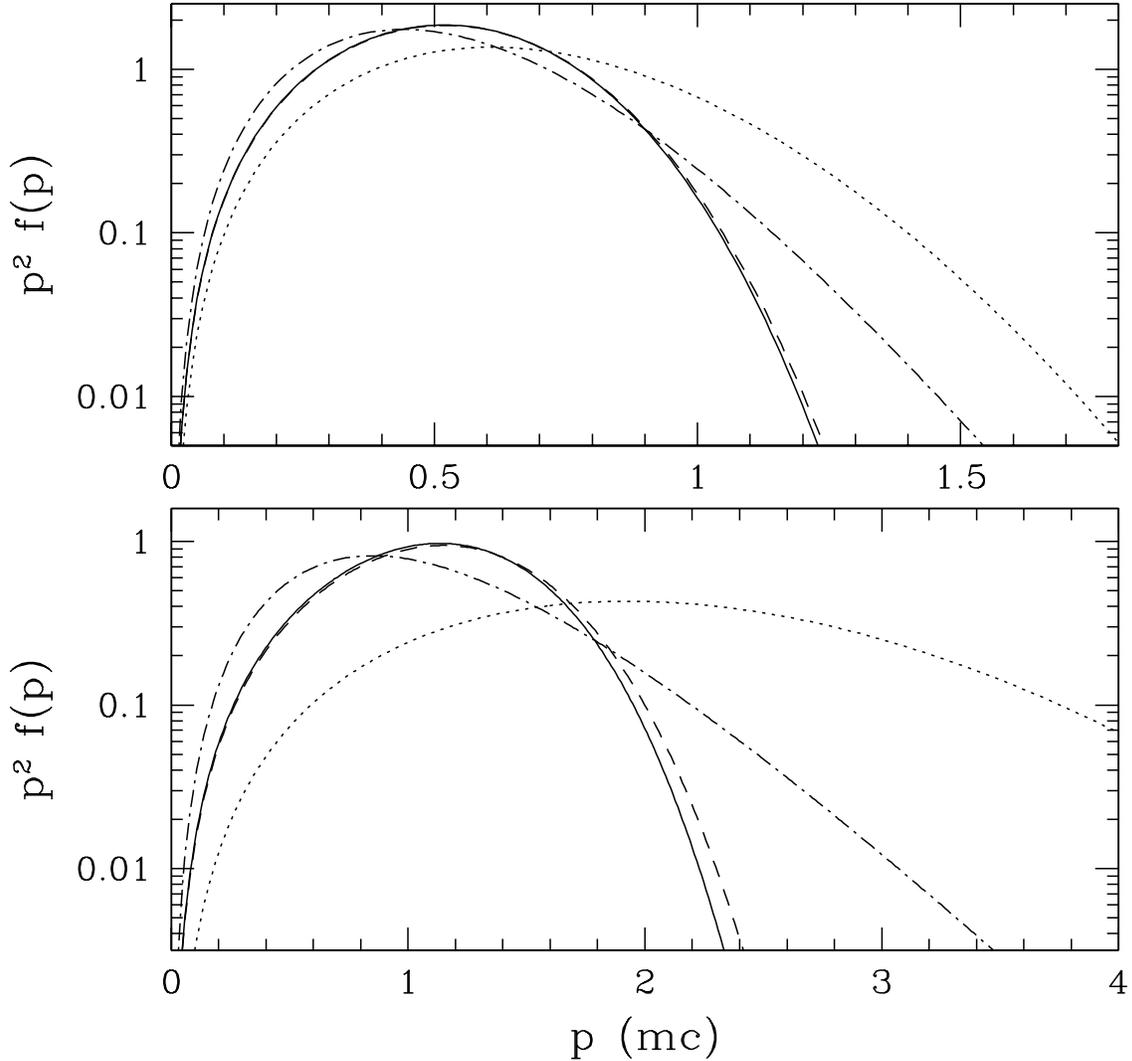}
\caption{(a) Equilibrium distribution of electrons located in
an isotropic field of self-absorbed synchrotron radiation with
$T_b^{\rm max}=5\cdot 10^{11}$~K. The solid line --- the exact result,
obtained using Eqs.~(\ref{feq_illar}), (\ref{d_illar}) and
(\ref{fsp}). The dotted line --- the nonrelativistic Maxwellian
distribution with $kT_{\rm eq}^0=95$~keV, a value found from
Eq.~(\ref{teq}). The dashed line (almost coincident with the
solid line) --- the mildly-relativistic approximation
(Eq.~[\ref{feq_fit}]). The dash-dotted line --- the relativistic
Maxwellian distribution with $kT=kT_{\rm eff}=45$~keV, a 
value found from Eq.~(\ref{teff}). The mean electron energy for this
distribution is equal to that for the equilibrium distribution. (b)
The same as (a), but $T_b^{\rm max}=5\cdot 10^{12}$~K, in which
case $kT_{\rm eq}^0=950~$~keV and $kT_{\rm eff}=140$~keV.
}
\label{fig_feq}
\end{figure}

\begin{figure}
\epsfxsize=17cm
\epsffile{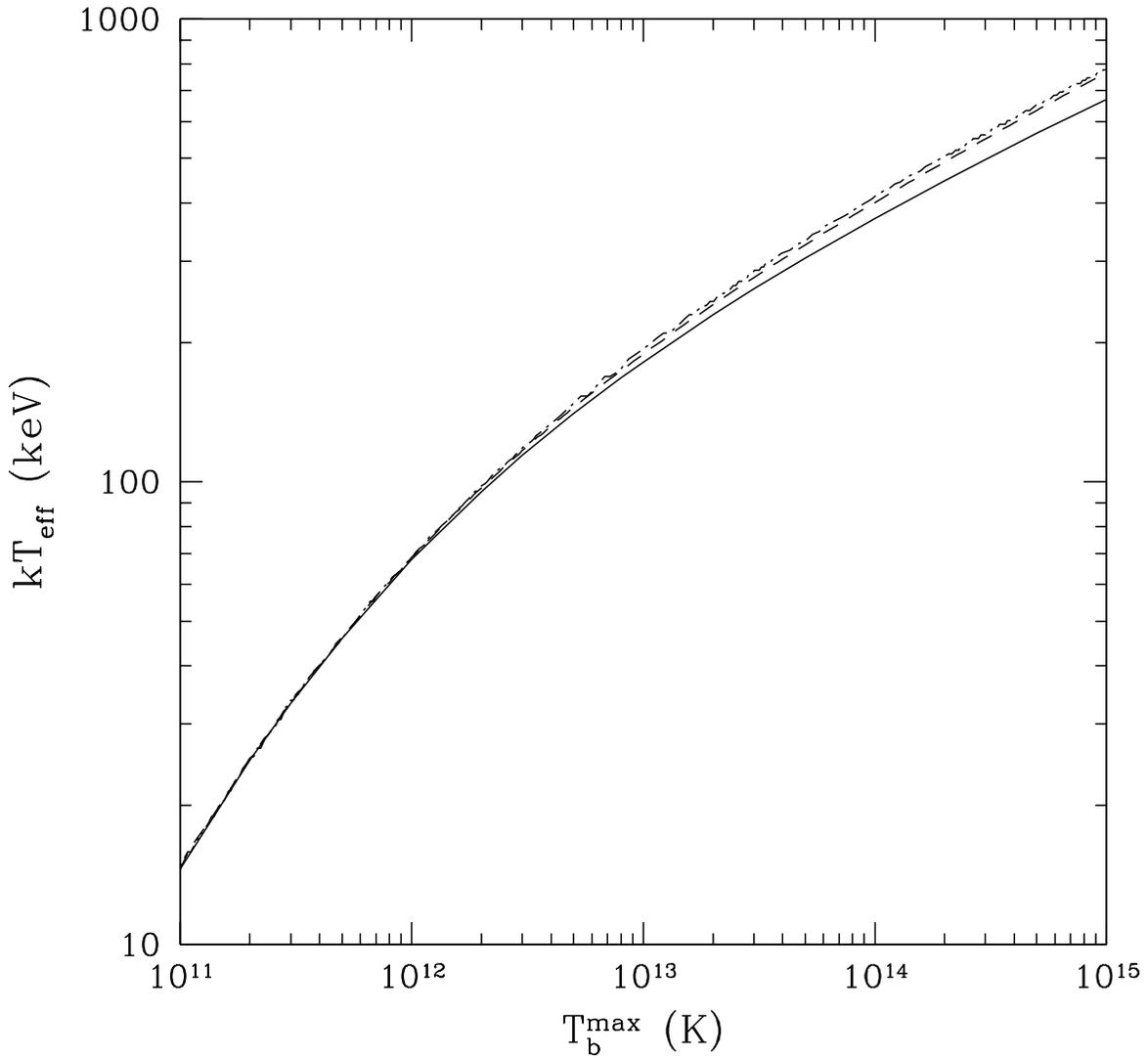}
\caption{Effective temperature of the stationary electron 
distribution, defined by Eq.~(\ref{teff}), vs. the peak radiation
brightness temperature. The solid line --- the exact result, obtained
using Eqs.~(\ref{feq_illar}), (\ref{d_illar}) and (\ref{fsp}). The
dashed line --- the result for the distribution given by the approximate
formula~(\ref{feq_p2}). The dependence of the stationary temperature
of Maxwellian electrons on $T_b^{\rm max}$ is reproduced from
Fig.~\ref{fig_teq} for comparison (the dash-dotted line). 
}
\label{fig_teff}
\end{figure}

\begin{figure}
\epsfxsize=17cm
\epsffile{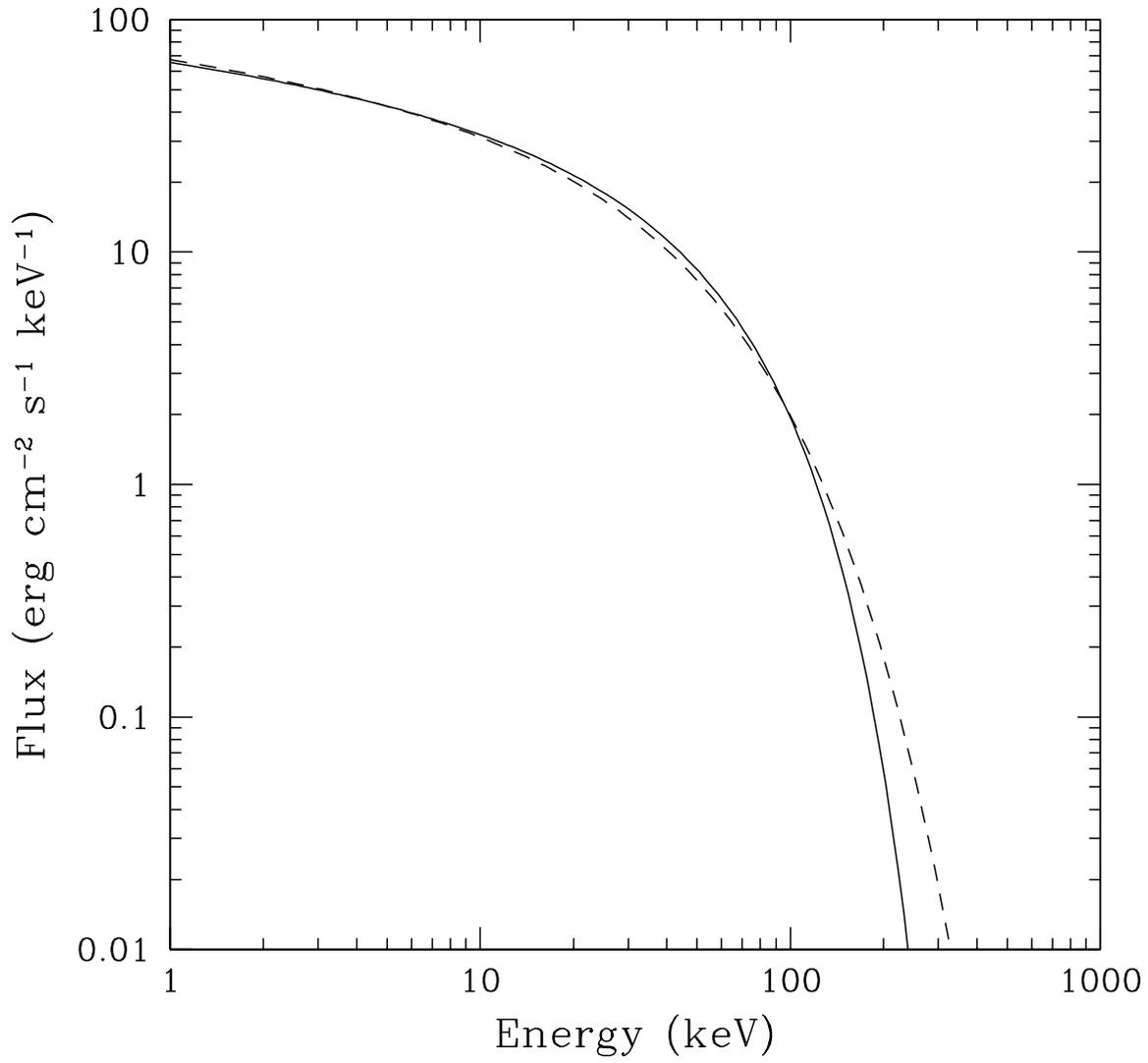}
\caption{Spectrum of hard X-ray bremsstrahlung emission from a plasma whose
electrons are in the distribution plotted in Fig.~\ref{fig_feq}a
(the solid line), compared with the spectrum (the dashed line) that
corresponds to the relativistic Maxwellian distribution with
$kT=kT_{\rm eff}=45~$~keV.
}
\label{fig_bremss}
\end{figure}

\end{document}